\documentclass[fleqn,twoside]{article}

\usepackage[headings]{espcrc2}

\readRCS
$Id: espcrc2.tex,v 1.2 2004/02/24 11:22:11 spepping Exp $
\usepackage{graphicx,epsfig}

\title{Detection of two-photon exclusive production of supersymmetric pairs at the LHC}

\author{Nicolas Schul\address[UCL]{Universit\'e catholique de Louvain, Center for Particle Physics and Phenomenology (CP3),\\ Louvain-la-Neuve, Belgium}
                        \thanks{e-mail: Nicolas.Schul@uclouvain.be} and
        Krzysztof Piotrzkowski\addressmark[UCL]}

\runtitle{Detection of $\gamma\gamma$ exclusive production of SUSY pairs}
\runauthor{N. Schul and K. Piotrzkowski}

\begin{document}

\begin{abstract}
The detection of pairs of sleptons, charginos and charged higgs bosons produced via photon-photon 
fusion at the LHC is studied, assuming a couple of benchmark points of the MSSM model. 
Due to low cross sections, it requires large integrated luminosity, but thanks to the striking signature of these
exclusive processes the backgrounds are low, and are well known. Very forward proton detectors can be used to measure the photon energies,
allowing for direct determination of masses of the lightest SUSY particle, of selectrons and smuons with a few GeV resolution.
Finally, the detection and mass measurement of quasi-stable particles predicted by the so-called sweet spot supersymmetry is discussed. 

\vspace{1pc}
\end{abstract}

\maketitle

%
%
\section{The two-photon exclusive pair production and search for new physics}
The $\gamma\gamma$ production of pairs of charged massive particles offers 
interesting potential for searches of the beyond standard model (BSM) particles. In a recent paper 
\cite{paper}, the initial comprehensive studies of high energy photon interactions at the LHC were 
reported. In the present contribution, the selected results discussed in ref. \cite{paper} are introduced 
and supplemented by new results. The exclusive two-photon production, $pp\rightarrow pXp$, provides clean experimental
conditions, thanks to absence of the proton remnants. Well defined final states can be then selected, and precisely reconstructed. 
Moreover, detection of the two final state protons, scattered at almost zero-degree angle, 
in the dedicated very forward detectors (VFDs), provides another striking signature, effective also at high
luminosity and with large event pile-up \cite{piotr,fp420}. In addition, the photon energies can be then measured and
used for the event kinematics recontruction (see section \ref{sec_vfd}). \\ 

The two-photon cross sections of pair production in general are determined by the mass, spin and charge of 
the produced particles, so the rate of produced particles at the LHC can be well predicted 
using the Equivalent Photon Approximation for the photon fluxes \cite{budnev}. One needs however to apply
corrections due to a possibility of strong interactions between protons, the so-called rescattering effects. 
The resulting suppression of the cross sections depends in particular on the invariant mass of the exclusively 
produced state $X$, and for the processes discussed in the following, as $pp\rightarrow pWWp$, it is estimated to be
about 10\% \cite{khoze}. This correction is ignored in the present analysis.
In table \ref{tab_Xsection}, the two-photon production cross sections computed using MadGraph/MadEvent \cite{madgraph}
for pairs of $W$ bosons (spin~$1$), and of fermions (spin~$1/2$) and scalars (spin~$0$) for a couple
of masses are shown \cite{paper}. 
\begin{table}[!ht]
\begin{center}
\caption{\small{Cross sections for several examples of the exclusive two-photon pair production at the LHC.
($F$ for fermion, $S$ for scalar). \cite{paper}}}
\label{tab_Xsection}
\begin{tabular}[!h]{lc||c}
\hline
Produced pairs & mass [GeV] & $\sigma$ [fb]  \\\hline
 $W^+ W^-$ & 80  & 108.5 \\
 $F^+ F^-$ & 100 & 4.064  \\
 $F^+ F^-$ & 200 & 0.399  \\
 $S^+ S^-$ & 100 & 0.680  \\
 $S^+ S^-$ & 200 & 0.069  \\
\hline
\end{tabular}
\end{center}
\end{table} 
One should note that if the inelastic production is considered,
when one of the protons dissociates into a low mass state, the cross-sections increase by about a factor of three
\cite{piotr}. \\

Supersymmetry predicts new scalar and fermionic particles above the W mass scale, and is therefore a prefect 
candidate for such novel and complementary searches using photon-photon collisions at the LHC. Two photon production
of non strongly-interacting SUSY particles at the LHC was discussed in \cite{zerwas}. In the present work, the 
experimental aspects of detection of such events at the LHC are studied. One should stress here, that in general 
the diffractive exclusive production of non strongly-interacting particles is heavily suppressed and can be safely
neglected\footnote{For example, the gluon mediated exclusive production of W boson pairs, is about 100 times smaller 
than the two-photon production at the LHC \cite{khoze}.}.
%
%
\section{Detection of supersymmetric pairs}
\begin{figure}[h!!]
\begin{center}
\includegraphics[scale=0.27]{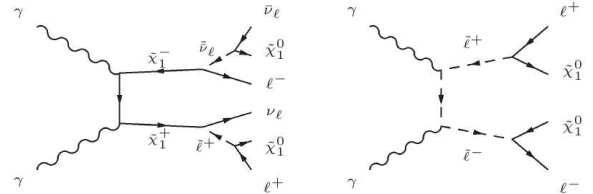}
\caption{\small{Most relevant Feynman diagrams for the MSSM analysis.
Left: Chargino decay into a charged(neutral) scalar and a neutral(charged) fermion.
Right: Slepton decay into a charged lepton and a LSP.}}
\label{fig_feynman}
\end{center}
\end{figure}

In two-photon processes, in general the production and decays mechanisms are simple, therefore not involving cascade decay
problems present in many SUSY studies at the LHC. In order to ensure the cleanest event signature
only fully leptonic final states are analyzed in the following. In the Minimal Supersymmetric Standard Model 
(MSSM) the lightest supersymmetric particle (LSP) is the stable neutralino $\tilde{\chi}_1^0$. As it is shown in figure 
\ref{fig_feynman}, the final state can be then composed of a pair of leptons and at least two
neutralinos which escape the detection, together with two neutrinos for some decays modes.\\

The requested final state for this MSSM search involves then: 
\begin{itemize}
	\item 2 leptons (ee or e$\mu$ or $\mu\mu$) of opposite charge detected in the central region,
	\item 2 scattered protons detected in the VFDs (assuming the tagged photon energy range: 20--900 GeV),
	\item missing energy due to escape of neutrinos and neutralinos,
	\item acoplanarity.
\end{itemize}
The only irreducible background process for this event topology is the exclusive two-photon production of 
pairs of $W$ bosons, followed by leptonic decays.
Indeed, $e^+e^-$, $\mu^+\mu^-$ and $\tau^+\tau^-$ pairs produced via $\gamma\gamma$ fusion can be easily 
supressed by requiring large acoplanarity and/or large missing energy. The cross section for the two-photon produced $WW$ pairs decaying leptonically 
reaches almost $7$ fb. \\

In sections \ref{LM1_section} and \ref{LM9_section}, the analysis is presented of two low mass benchmark points in the usual MSSM plane 
(LM1 and LM9) \cite{benchmark}, chosen to study the potential for detection of SUSY particles and of determination of their masses. 
For LM1 and LM9, respectively, the slepton and chargino contributions are dominant. 
Third scenario discussed in section \ref{sweet_section} involves the $\tilde{\tau}_1^\pm$ as the NLSP, and 
the analysis method described above does not apply. Signal as well as background samples were generated
using the modified CalcHEP \cite{calchep} and then passed to the modified Pythia \cite{pythia}, 
where the decay and hadronisation steps were performed. In order to simulate acceptance regions of the
LHC central detectors, cuts (\ref{cut}) were applied on leptons at generator level:
\begin{eqnarray}
\label{cut}
&p_T(e^\pm) > 10~\textrm{GeV}, \ \ \ \ p_T(\mu^\pm) > 7~\textrm{GeV}& \nonumber \\ 
&|\eta(\ell^\pm)| < 2.5 .&
\end{eqnarray}

%
%
\section{LM1 benchmark scenario}
\label{LM1_section}
Sparticle masses in the LM1 scenario are given in table \ref{tab_mass_1}. Application of the acceptance cuts (1) 
reduces the cross sections by factors between 1.5 and 50, depending on the particle type.\\ 
\begin{table}[ht!]
\begin{center}
\caption{\small{Masses of LM1 MSSM particles derived from running the RGE for $m_0 = 60$~GeV, $m_{1/2} = 250$~GeV,
 tg($\beta$) = 10, $A_0$ = 0, $\mu > 0$. ($\ell$ = $e, \mu$)}}
\label{tab_mass_1}
\renewcommand{\arraystretch}{1.2} 
\begin{tabular}[!h]{cc||cc}
\hline
\multicolumn{4}{c}{mass [GeV]} \\
\hline
$\tilde{\ell}_R^{\pm}$ & 118 & $\tilde{\chi}_1^{\pm}$ & 178 \\
$\tilde{\ell}_L^{\pm}$ & 187 & $\tilde{\chi}_2^{\pm}$ & 360 \\
$\tilde{\tau}_1^{\pm}$ & 111 & $H^{\pm}$ & 381 \\
$\tilde{\tau}_2^{\pm}$ & 190 & $\tilde{\chi}_1^{0}$ & 96 \\
\hline
\end{tabular}
\end{center}
\end{table}

Because the branching ratio $BR(\tilde{\ell}^+~\to~\ell^+~+~\tilde{\chi}_1^0) \simeq 100\%$,
selectron and smuon pairs are the major expected contribution in the analysis of di-leptonic final states.
In addition, almost $65\%$ of right sleptons and $75\%$ of left ones fall within the acceptance window (\ref{cut}). 
On the contrary, pairs of charginos (left diagram of figure \ref{fig_feynman}) have other significant decay modes
(including hadronic decays) therefore only $7\%$ of the produced chargino pairs will be detected. 
Staus, since they decay into tau leptons, will produce mostly two $\tau$-jet final states, hence
only $1$ and $6\%$ of the $\tilde{\tau}_1$ and the $\tilde{\tau}_2$ pairs, pass acceptance cuts, respectively. 
One could also consider final states with one lepton and one $\tau$-jet or with two $\tau$-jets, but the statistical 
improvement is then only relevant for large tg($\beta$) models, where couplings to (s)taus are enhanced. 
Finally, no relevant signal of the charged higgs boson pairs can be seen, since 
$H^+ \to \bar{b}t$ is the dominant decay mode.
\subsection{Lepton flavour sharing}
By considering only the information from the central detector and applying acceptance cuts, one already reaches a 
signal-to background ($S/B$) ratio close to $1/5$, see table \ref{tab_sigma_cut_1b}. 
This ratio can be easily improved by considering only same flavour dileptonic events.
Indeed, because the signal is dominated by $\tilde{\ell}_R$ and $\tilde{\ell}_L$ pair decays, around $90\%$ of the
LM1 events is composed of $e^+e^-$ and $\mu^+\mu^-$ leptons (see right diagram of figure \ref{fig_feynman}). At the 
same time the background $W$ pairs are suppressed by a factor $2$ by selecting same flavour leptons. 
\subsection{Very forward detector information}
\label{sec_vfd}
In contrast to the nominal $pp$ studies which can only use kinematical quantities measured with the central tracker, 
calorimeters, etc., the main experimental advantage of the two-photon processes relies on the detection in VFDs of the two forward 
scattered protons \cite{piotr,xavier}. To evaluate impact of these detectors on the analysis, a complete set of 
VFDs, situated at $220$~m and $420$~m on each side of the interaction point, is assumed. In this case, the accepted or 
tagged photon energy is contained between 20 and 900 GeV.  The efficiency of detecting both forward protons is then around 
$80\%$ for the LM1 benchmark events, so almost whole relevant photon spectrum is probed. 
Finally, one should note that the proposed forward detectors are capable of running at high luminosity, 
allowing for selection of such exclusive dilepton events also in presence of large event pileup \cite{fp420}.\\
 
In order to simulate the photon energy resolution, as can be obtained using the VFDs \cite{hector}, 
a gaussian smearing of the photon energy with the following simple parametrization of its standard deviation:
\begin{center}
\begin{equation}
\sigma_{E_{\gamma}} = \textrm{max} (1.5~\textrm{GeV},~E_\gamma / 100) .
\end{equation}
\end{center}
Useful quantities used to discriminate signal and background are listed below. An obvious variable
which can be computed is the total photon-photon center-of-mass energy,
\begin{center}
\begin{equation}
W_{\gamma\gamma} = 2 \sqrt{E_{\gamma_1} E_{\gamma_2}}
\end{equation}
\end{center}
where $E_{\gamma_1}$, $E_{\gamma_2}$ are the reconstructed energies of two colliding photons. 
\begin{figure}[ht!]
\begin{center}
\includegraphics[scale=0.38]{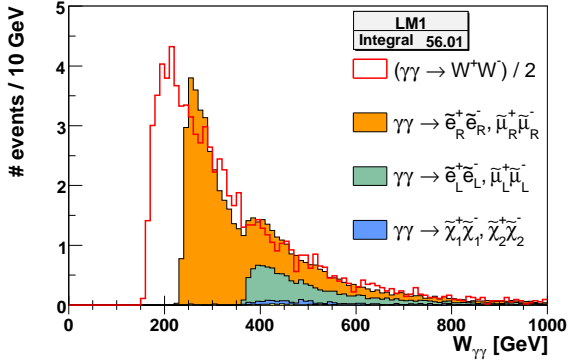}
\caption{\small{Distribution of two-photon invariant mass $W_{\gamma \gamma}$ for the LM1 benchmark and integrated luminosity
L = 100 fb$^{-1}$. Two visible peaks are due to production thresholds of $\tilde{\ell}_R^+\tilde{\ell}_R^-$ and 
$\tilde{\ell}_L^+\tilde{\ell}_L^-$ pairs. Verious contribution are added cumulatively. 
The background distribution of $WW$ pairs is shown separately, and is rescaled to obtain similar size as signal.}}
\label{fig_wgg1}
\end{center}
\end{figure}
The expected $W_{\gamma\gamma}$ distribution is shown in figure \ref{fig_wgg1} for an integrated luminosity of
100~fb$^{-1}$.
One can see two significant peaks due to the production thresholds of right slepton (around 250~GeV) and of left sleptons (around 400~GeV).
In this way, the slepton mass spectrum can be probed by measuring the threshold energy for each peak, which is approximately
equal to the sum of masses of the two produced sparticles. It should be stressed that the mass determination
in this method depends only on the VFD energy resolution, and not on the resolutions of the central detectors. 
Actual precision of $\tilde{\ell}_R^\pm$ and $\tilde{\ell}_L^\pm$ mass determination is then mostly driven by available 
statistics. Moreover, this quantity can also be used to suppress the background since the $W_{\gamma\gamma}$ distribution for 
$W$ pairs is well known and starts at about $2m_W$. The choice of the analysis cut $W_{\gamma\gamma}^{min}$ could be then 
changed according to the SUSY mass constraints coming from $pp$ studies.\\ 

The missing energy carried away by the neutrinos and the neutralinos can be estimated as
\begin{center}
\begin{equation}
E_{miss} = E_{\gamma_1} + E_{\gamma_2} - E_{l_1} - E_{l_2}
\end{equation}
\end{center}
where $E_{l1},~E_{l2}$ are the measured leptons energies. A conservative correction is made
to account for the \emph{bremsstrahlung} in electronic decays. It is assumed that the soft
bremssthralung photons, $p_T(\gamma) < 10$~GeV, are not detected. This results in the biased $E_{miss}$
in a small fraction of events, but otherwise leptons are very well reconstructed in the central detectors. 
Therefore, it is assumed that the energy and momentum of the leptons are known exactly, and the resolution of the
reconstructed kinematical variables is dominated by the photon energy resolutions.\\

\begin{figure}[ht!]
\begin{center}
\includegraphics[scale=0.38]{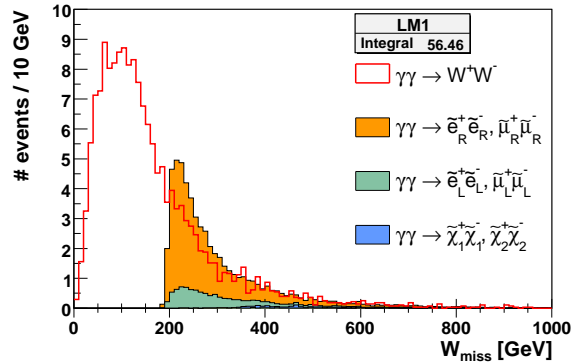}
\caption{\small{Distribution of missing invariant mass $W_{miss}$ for the LM1 MSSM benchmark for the integrated luminosity
L = 100 fb$^{-1}$. It starts at about $2~m_{LSP}$ for SUSY, at zero for the $WW$ background.}}
\label{fig_wmiss1}
\end{center}
\end{figure}

The missing invariant mass distribution can then be defined as
\begin{center}
\begin{equation}
W_{miss} = \sqrt{E_{miss}^2 - P_{miss}^2}
\end{equation}
\end{center}
where $P_{miss}$ is the event missing momentum and is calculated in analogy to $E_{miss}$. Missing
mass is on average larger for the SUSY event sample since a supersymmetric event will always produce at least
two massive LSPs. This can be seen in figure \ref{fig_wmiss1} where the expected distributions of 
$W_{miss}$ for 100~fb$^{-1}$ is shown. 
SUSY distribution is peaked slightly above $200$~GeV, which is twice the mass of
the lightest neutralino in this model, while it starts at zero for the SM contribution (which is not rescaled here). 
This quantity allows for measuring the LSP mass with a high resolution and to suppress the $W^+W^-$ background by 
requiring a $W^{min}_{miss}$ cut.
\subsection{Estimate of the LM1 signal significance}
\begin{figure}[ht!]
\begin{center}
\includegraphics[scale=0.38]{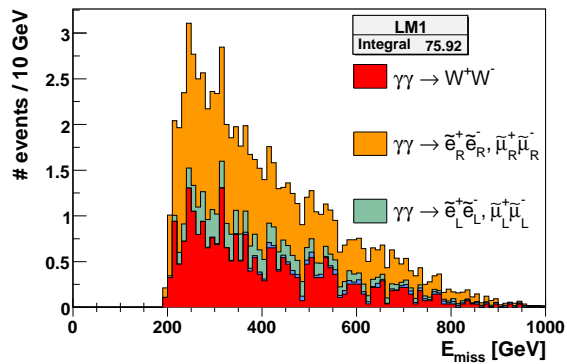}
\caption{\small{Missing energy distribution for the LM1 benchmark and the $WW$ background for the integrated
luminosity L = 100 fb$^{-1}$.}}
\label{fig_emissex}
\end{center}
\end{figure}

To reconstruct $W_{\gamma\gamma}$ and $W_{miss}$ quantities, it is assumed that the both
forward scattered protons are detected in the VFDs within the full tagging range. Then, by applying the following analysis cuts
\begin{itemize}
\item acceptance cuts (\ref{cut}),
\item $W_{\gamma\gamma} >$ 200~GeV and $W_{miss} >$ 186~GeV,
\item lepton angular cuts, $\Delta \eta < 2.1$ and $\Delta R < 3.0$ ($R$ is the distance between two leptons on the pseudorapidity--phi angle plane),
\item same flavour lepton selection,  
\end{itemize}
one reaches a $S/B$ ratio close to $2$. The various contributions for the signal are  
given in table \ref{tab_sigma_cut_1b} and shown in figure \ref{fig_emissex} for the integrated luminosity L = 100~fb$^{-1}$, 
and $S \simeq 51$ and $B \simeq 26$ events.
\begin{table}[ht!]
\begin{center}
\caption{\small{LM1 signal and $WW$ background cross sections before and after the acceptance cuts (including the flavor selection), 
and after the analysis cuts. 
Values are given in fb. ($\ell~=~e$,~$\mu$. \ \ $i= 1, 2$).}}
\label{tab_sigma_cut_1b}
\renewcommand{\arraystretch}{1.2} 
\begin{tabular}[!h]{l||c|c|c}
\hline
Processes & $\sigma$ & $\sigma_{acc}$ & $\sigma_{acc + ana}$ \\
\hline
$\gamma\gamma\to\tilde{\ell}_R^+\tilde{\ell}_R^-$ & 0.798 & 0.522 & 0.403 \\
$\gamma\gamma\to\tilde{\ell}_L^+\tilde{\ell}_L^-$ & 0.183 & 0.135 & 0.089 \\
$\gamma\gamma\to\tilde{\tau}_i^+\tilde{\tau}_i^-$ & 0.604 & 0.054 & 0.003 \\
$\gamma\gamma\to\tilde{\chi}_i^+\tilde{\chi}_i^-$ & 0.642 & 0.043 & 0.014 \\
$\gamma\gamma\to H^+H^- $                         & 0.004 & /     & / \\
                                                  &       &       &   \\
$\gamma\gamma\to W^+W^- $                         & 108.5 & 3.820 & 0.255 \\
\hline
\end{tabular}
\end{center}
\end{table}

The $5 \sigma$ discovery for the LM1 left and right sleptons is then reached already after 25~fb$^{-1}$ 
thanks to strong suppression of the irreducible background. It could still be improved by
using additional cuts exploring the correlation between $W_{\gamma\gamma}$ and $W_{miss}$ as it is done for the 
LM9 study below. Finally, it could be improved even further by including the inelastic two-photon production, in this 
case however only one proton is detected and the kinematical reconstruction is not so effective.
For the same benchmark point, the nominal proton-proton studies claim  $5 \sigma$ discovery
after about 10~fb$^{-1}$\cite{cmsnote}. 
However, determination of sparticle masses in this case is much more complicated.\\ 

Similar two-photon analyses can be done for other benchmark points with low slepton masses as LM2, LM4 and LM6.

\subsection{Mass measurement}
The main advantage of the two-photon analysis is large sensitivity to sparticle masses. Mass determination using 
the production threshold values in $W_{\gamma\gamma}$ and $W_{miss}$ distributions is limited by the number of selected events. 
Another approach, based on other kinematic quantities, can provide a method to measure mass of the sleptons more
on the event-by-event basis.\\

\begin{figure}[ht!]
\begin{center}
\begin{tabular}{cc}
\epsfig{file=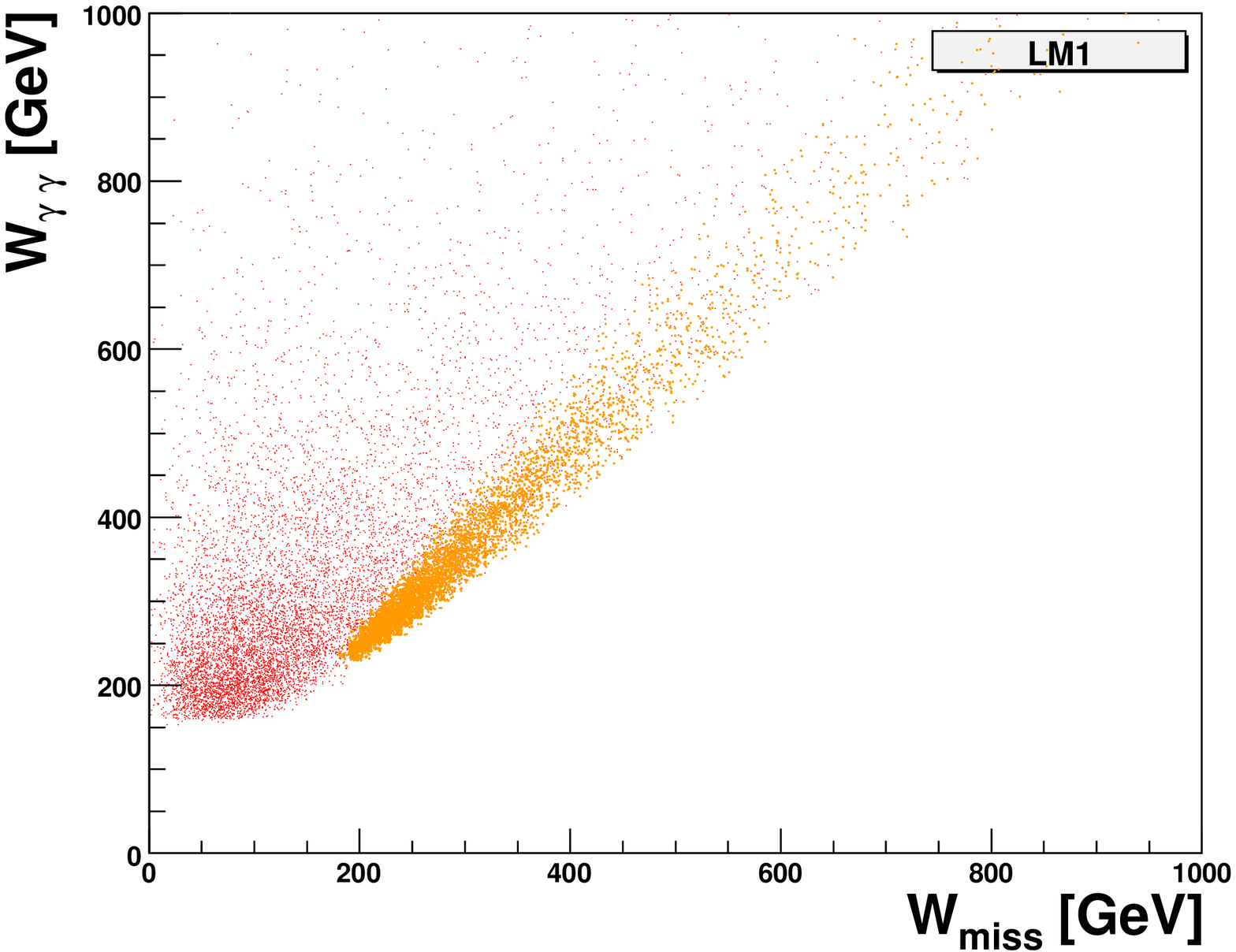, width=7.0cm,clip} \\
\epsfig{file=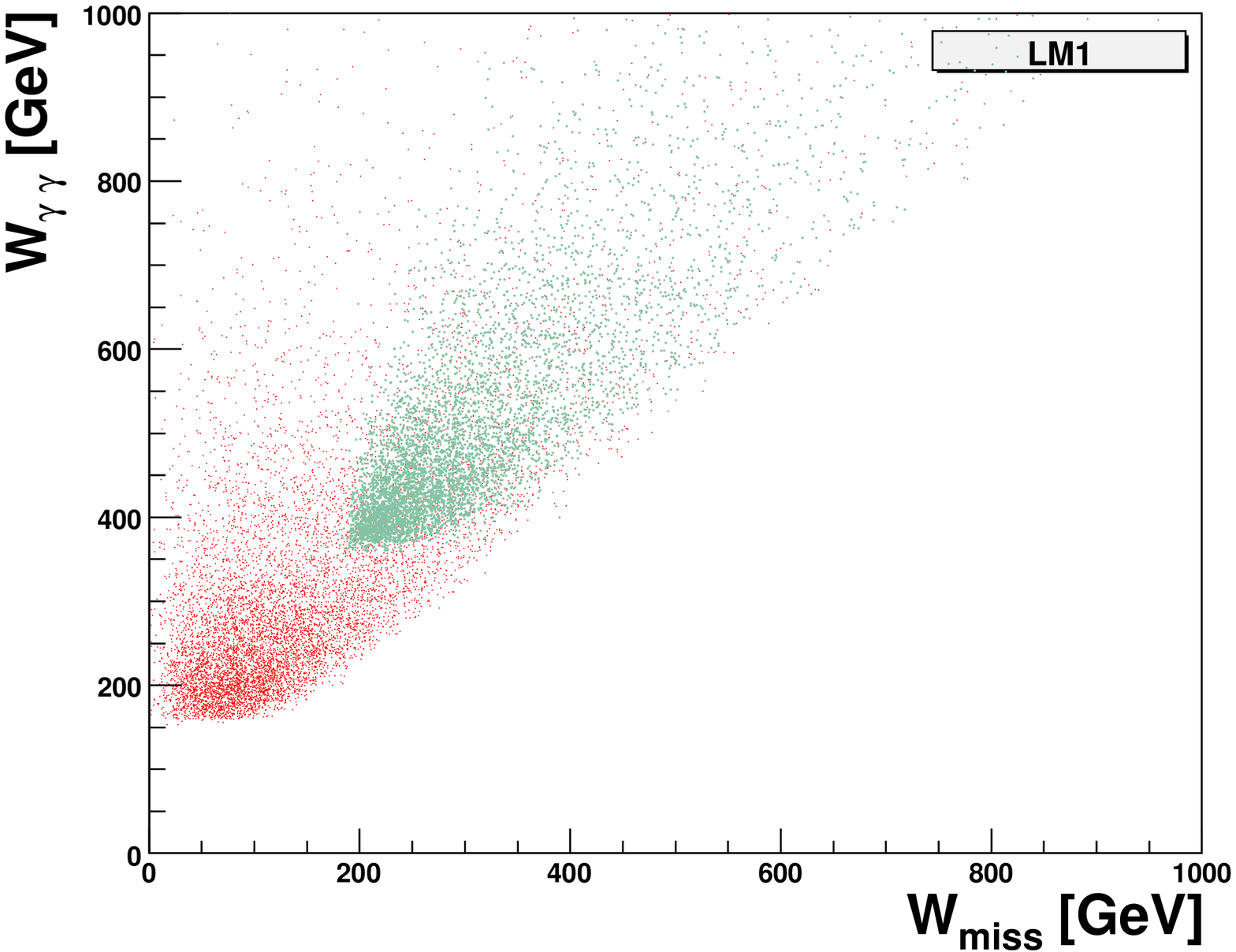, width=7.0cm,clip} 
\end{tabular}
\caption{\small{Scatter plots for the LM1 signal and the $WW$ background on the $W_{miss}$, $W_{\gamma\gamma}$ plane. 
Upper: $\tilde{\mu}_R^+\tilde{\mu}_R^-$ events (orange/light) and $W^+W^-$ events (red/dark). 
Lower: $\tilde{\mu}_L^+\tilde{\mu}_L^-$ events (green/light) and $W^+W^-$ events (red/dark).}}
\label{fig_2D}
\end{center}
\end{figure}

The two-dimensional plots at figure \ref{fig_2D} represent event distributions on the $W_{\gamma\gamma}$, 
$W_{miss}$ plane for the MSSM processes and for the $W^+W^-$ background, after acceptance cuts. One can 
observe that for the signal events these two variables are strongly correlated, and much less so for the background.
Moreover, it can be shown that the width of the distribution is related to the mass of the produced sparticles. 
This demonstrates a close relationship between the MSSM masses, the $\gamma\gamma$ invariant mass and the missing mass. An empirical
quantity has been built in order to take into account this observation:
\begin{flushleft}
$(2m_{reco})^2 = W_{\gamma \gamma}^2 - ( [W_{miss}^2 - 4 m_{\tilde{\chi}_1^0}^2]^{1/2}$
\end{flushleft}
\begin{flushright}
$ + [W_{lep}^2 - 4 m_{lep}^2]^{1/2} )^2$
\end{flushright}
where $W_{lep}$ is the invariant mass of the two lepton system, and $m_{lep}$ is the lepton mass, and $2m_{reco}$ 
is the reconstruted mass of the produced sparticles. That relation does not work well for 
$\tilde{\tau}$ and $\tilde{\chi}$ pairs because in general they decay into final states with 
more neutrinos and neutralinos. The only needed input in this method is the value of the LSP mass, 
which can be derived from the $W_{miss}$ distribution.\\

\begin{figure}[ht!]
\begin{center}
\includegraphics[scale=0.38]{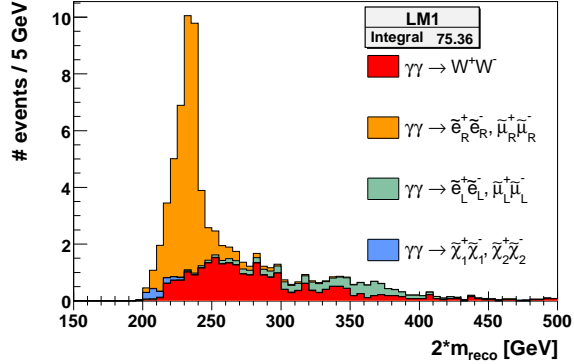}
\caption{\small{Cumulative distributions of the reconstructed mass $2m_{reco}$ for the LM1 signal and the $WW$ background for 
the intergrated luminosity L = 100 fb$^{-1}$.}}
\label{fig_mass}
\end{center}
\end{figure}

The reconstruction power of this empirical quantity is illustrated in figure \ref{fig_mass} for the integrated luminosity 
L = 100~fb$^{-1}$. A narrow peak centered at $2m_{reco} = $ 236~GeV = 2~$\times$~118~GeV, allows for efficient and direct
determination of the $\tilde{e}_R^\pm$ and $\tilde{\mu}_R^\pm$ mass. A second peak, centered at $2m_{reco} = $ 370~GeV = 2~$\times$~187~GeV, 
with larger width, corresponds to $\tilde{e}_L^\pm$ and $\tilde{\mu}_L^\pm$ pairs but is not so well visible. Right
selectron and smuon mass might be determined using this empirical method with a few GeV resolution.
In summary, this techniques of reconstruction of sparticles masses are very promising for the low SUSY mass scenarios, 
and it should be further studied using more complete, or even full simulations of the central detector performance.
%
%
\section{LM9 benchmark scenario}
\label{LM9_section}
In this second scenario, photon-photon processes would probe the MSSM signal via the detection of light chargino
pairs. In LM9, in order to keep the mass for the lightest neutral higgs above the LEP limit, scalar sparticles are very 
massive (around $1.5$~TeV, see table \ref{tab_mass_2}) and consequently can not be detected using the two-photon events. 
\begin{table}[ht!]
\begin{center}
\caption{\small{Masses of the LM9 MSSM particles derived from running the RGE for $m_0 = 1450$~GeV, $m_{1/2} = 175$~GeV,
 tg($\beta$) = 50, $A_0$ = 0, $\mu > 0$. ($\ell$ = $e, \mu$)}}
\label{tab_mass_2}
\renewcommand{\arraystretch}{1.2} 
\begin{tabular}[!h]{cc||cc}
\hline
\multicolumn{4}{c}{mass [GeV]} \\
\hline
$\tilde{\ell}_R^{\pm}$ & 1450 & $\tilde{\chi}_1^{\pm}$ & 107 \\
$\tilde{\ell}_L^{\pm}$ & 1450 & $\tilde{\chi}_2^{\pm}$ & 223 \\
$\tilde{\tau}_1^{\pm}$ & 1054 & $H^{\pm}$ & 495 \\
$\tilde{\tau}_2^{\pm}$ & 1267 & $\tilde{\chi}_1^{0}$ & 65 \\
\hline
\end{tabular}
\end{center}
\end{table} 
The only significant contribution for the LM9 signal arises from the lightest chargino detection. In that 
particular scenario, the branching ratio $BR(\tilde{\chi}_1^+~\to~\tilde{\chi}_1^0~+~W^{*+})$~=~100$\%$, and the signal and
background decay schema are almost the same, with additional missing energy in SUSY events. After applying only the 
acceptance cuts (\ref{cut}) on generator level, one finds for the LM9 point a $S/B$ ratio of about $2.5\%$
 with about 10 $\tilde{\chi}_1^+\tilde{\chi}_1^-$ events for the integrated luminosity L~=~100~fb$^{-1}$ 
(see first columns of table \ref{tab_LM9}).\\

\begin{figure}[ht!]
\begin{center}
\includegraphics[scale=0.33]{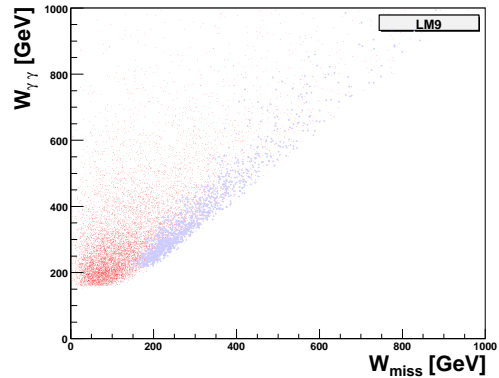}
\caption{\small{Scatter plots for the LM9 signal and $WW$ background events on the $W_{miss}$, $W_{\gamma\gamma}$ plane.}}
\label{fig_LM9_2D}
\end{center}
\end{figure}

As for the LM1 study, one can also apply analysis cuts on $W_{\gamma\gamma}$ and $W_{miss}$ quantities 
as it can be deduced from figure \ref{fig_LM9_2D}. Best values were found to be:
\begin{itemize}
\item $W_{\gamma\gamma} > 220$~GeV and $W_{miss} > 180$~GeV,
\item $W_{\gamma\gamma} < 1.5~W_{miss}$,
\item $W_{lep} < 170$~GeV,
\item no flavour selection,
\item angular cuts $\Delta R < $2.6 and $\Delta\eta < $3.5 ($R$ is the distance between two leptons in the pseudorapidity--phi angle plane),
\end{itemize}
Additional cuts, in comparison to the LM1 study, the cut on two-lepton invariant mass and the cut using the correlation 
between $W_{\gamma\gamma}$ and $W_{miss}$ are applied in order to suppress two-photon $WW$ events, in situation where the flavor
selection is not effective. 
The expected visible cross sections after applying these analysis cuts are quoted 
in table \ref{tab_LM9}, and result in respectively $7$ and $41$ events for the MSSM signal and the $WW$ background for 100~fb$^{-1}$.
\begin{table}[ht!]
\begin{center}
\caption{\small{Cross sections before and after the acceptance cuts, and after the analysis cuts for the LM9 signal
and the $WW$ background. Values are given in fb.}}
\label{tab_LM9}
\renewcommand{\arraystretch}{1.2} 
\begin{tabular}[!h]{l||c|c|c}
\hline
Processes & $\sigma$ & $\sigma_{acc}$ & $\sigma_{acc + ana}$ \\
\hline
$\gamma\gamma\to\tilde{\chi}_1^+\tilde{\chi}_1^-$ & 3.281 & 0.099 & 0.071 \\
$\gamma\gamma\to\tilde{\chi}_2^+\tilde{\chi}_2^-$ & 0.273 & 0.001 & / \\
                                                  &       &       &   \\
$\gamma\gamma\to W^+W^- $                         & 108.5 & 3.804 & 0.408 \\
\hline
\end{tabular}
\end{center}
\end{table}
Those stringent cuts improve the $S/B$ ratio to above $15\%$, without decreasing dramatically the number of signal
events. Another analysis method taking into account semi-leptonic and fully-hadronic final states should
then be considered in order to tackle these particular supersymmetric scenarios involving chargino 
pairs as dominant contribution. Consequently, the event signature involving jets and $\tau$-jets would be more
difficult to extract.
%
%
\section{Sweet Spot SUSY scenario}
\label{sweet_section}
In this last scenario, SUSY search strategy would be changed due to different phenomenology. 
In some non-common SUSY theories, the LSP is not $\tilde{\chi}_1^0$ as in MSSM but rather the gravitino.
One interesting framework in which that occurs is the \emph{Sweet Spot SUSY} \cite{sweet}, where the 
next-to-LSP is predicted to be the lightest tau $\tilde{\tau}_1^+$, with low mass around $116$~GeV. It would
be quasi-stable, with a decay time of $\cal{O}$(3000s). Production of $\tilde{\tau}_1^{\pm}$ pairs in 
photon-photon interactions will then be detected by observing very exclusive final states -- two heavy and stable, opposite 
charge particles produced centrally plus two forward scattered protons. Using the phenomenological spectrum of \cite{sweet}, 
one finds the cross section for the two-photon $\tilde{\tau}_1^+$ production of about $0.43$~fb.\\

\begin{figure}[ht!]
\begin{center}
\includegraphics[scale=0.35]{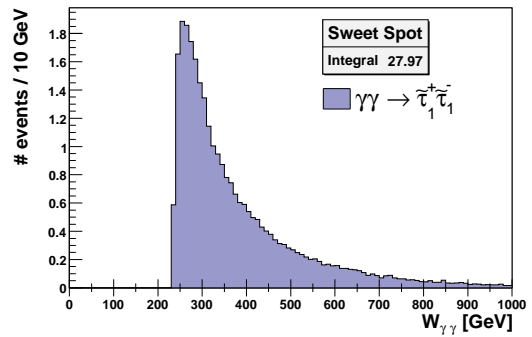}
\caption{\small{Photon-photon invariant mass distribution after the integrated luminosity L~=~100~fb$^{-1}$for 
$pp(\gamma\gamma\to\tilde{\tau}_1^+\tilde{\tau}_1^-)pp$ with
$p_T(\tilde{\tau}_1^\pm) > 10$~GeV requirement.}}
\label{fig_sweet}
\end{center}
\end{figure}

A usual method in $pp$ studies to detect such heavy lepton like pairs relies on the use of the $dE/dx$ variable, or
time of flight measurements, and results in a poor reconstruction using calorimeters and muon chambers \cite{HSCP}. 
On the contrary, in the two-photon analysis, almost all the kinematic information is available, again thanks to the
detection of the forward scattered protons. For example, the two-photon invariant mass $W_{\gamma\gamma}$ as shown in figure \ref{fig_sweet} 
for the integrated luminosity L~=~100~fb$^{-1}$, assuming detection if both staus with $p_T(\tilde{\tau}_1^\pm) > 10$~GeV, and of two 
forward protons in the full VFD setup. The stau mass can then be directly measured by comparing the two-photon invariant mass with the 
stau momenta. Such a event-by-event mass measurement has a very good resolution, of about 5~GeV.
Finally, the stau spin could be determined by analyzing the stau angular distributions in their center-of-mass reference system.

%
%
\section{Conclusion}
Photon-photon interactions provides an additional and complementary handle on the BSM theories in general, and on the supersymmetric low mass models 
in particular. As it has been shown, the main advantage of the two-photon production analyses lies in the extra kinematical information --
as the total center of mass energy, the missing energy, or the pair boost -- obtained using the very forward detectors (at 220~m and 420~m) 
at the LHC, as well as in the cleanliness of the event signature.\\

Discovery of the left and right-handed selectrons and smuons after L~=~25~fb$^{-1}$ for the light MSSM scenarios is expected and mass measurement 
of these sparticles as well as of the lightest SUSY particle with a few GeV resolution should be possible. This assumes 
high resolution of the VFDs, also during high luminosity runs.


\begin{thebibliography}{99}
\bibitem{paper} J. de Favereau \emph{et al.}, {\it \small High energy photon interactions at the LHC}, CP3--08--04, June 2008, to be submitted to EPJC.
\bibitem{piotr} K.~Piotrzkowski, {\it \small Tagging two-photon production at the LHC}, Phys. Rev. D63 (2001) 071502.
\bibitem{fp420} M. Albrow \emph{et al.}, {\it \small The FP420 R$\&$D Project: Higgs and New Physics with forward protons at the LHC FP420},
arXiv:0806.0302 [hep-ex].
\bibitem{budnev} V. M. Budnev \emph{et al.}, {\it \small The Two photon particle production mechanism. Physical problems. Applications. Equivalent photon approximation.},
Phys.~Rept.~15 (1974) 181.
\bibitem{khoze} B. E. Cox \emph{et al.}, {\it \small Detecting the standard model Higgs boson in the WW decay channel using forward proton tagging at the LHC}, Eur. Phys. J. C45 (2006) 401.
\bibitem{madgraph} J. Alwall \emph{et al.}, {\it \small MadGraph/MadEvent v4: The New Web Generation}, JHEP 0709:028 (2007).
\bibitem{zerwas} J.~Ohnemus, T~Walsh, P.~Zerwas, {\it \small $\gamma\gamma$ production of non--strongly interacting SUSY particles at hadron colliders}, Phys. Lett. B328 (1994) 369.
\bibitem{benchmark} M.~Battaglia \emph{et al.}, {\it \small Updated post-WMAP benchmarks for supersymmetry}, Eur. Phys. J. C33 (2004) 273.
\bibitem{calchep} A.~Pukhov, {\it \small CalcHEP}, Nucl. Inst. Meth A502 (2003) 596.
\bibitem{pythia} T.~Sj\"ostrand \emph{et al.}, {\it \small PYTHIA 6.4}, Comput. Phys. Commun. 135 (2001) 238. 
\bibitem{xavier} X.~Rouby, {\it \small Tagging photon interactions at the LHC}, these proceedings. 
\bibitem{hector} J. de Favereau de Jeneret, X. Rouby and K. Piotrzkowski, {\it \small HECTOR: A Fast simulator for the transport of particles in beamlines}, JINST 2: P09005 (2007), arXiv:0707.1198v1 [physics.acc-ph] (CP3--07--13).
\bibitem{cmsnote} Y.~Andreev \emph{et al.}, {\it \small Using $\ell^+\ell^- + E_T$miss + jet veto signature for slepton detection}, CMS NOTE 2006/132
\bibitem{sweet} M.~Ibe, R.~Kitano, {\it \small Sweet Spot Supersymmetry}, JHEP 0708:16 (2007).
\bibitem{HSCP} M.~Fairbairn \emph{et al.}, {\it \small Stable Massive Particles at Colliders}, Phys. Rept. 438 (2007) 1.

\end{thebibliography}
\end{document}